\documentclass[twocolumn,floatfix,superscriptaddress,a4paper,showpacs,showkeys,nofootinbib,reprint,prc]{revtex4-1}

\usepackage[colorlinks=true,linktocpage=true,linkcolor=blue,citecolor=blue,allcolors=blue]{hyperref}
\usepackage{epsfig}
\usepackage{latexsym}
\usepackage{xspace}
\usepackage[utf8]{inputenc}
\usepackage{indentfirst}
\usepackage{enumerate}
\usepackage{color}
\usepackage{tabularx}

\usepackage{amsmath}
\usepackage{amssymb}
\usepackage[english]{babel}
\usepackage{url}
\newcommand{\eq}[1]{\begin{align} #1 \end{align}}

\newcommand{\trit}{t}
\newcommand{\tHe}{$^3$He}

\newcommand{\fHe}{$^4$He}

\begin{document}


\title{
Feeddown contributions from unstable nuclei\\
in relativistic heavy-ion collisions
}

\author{Volodymyr Vovchenko}
\affiliation{Nuclear Science Division, Lawrence Berkeley National Laboratory, 1 Cyclotron Road, Berkeley, CA 94720, USA}
\affiliation{
Institut f\"ur Theoretische Physik,
Goethe Universit\"at Frankfurt, Max-von-Laue-Str.\ 1, D-60438 Frankfurt am Main, Germany}
\affiliation{Frankfurt Institute for Advanced Studies, Giersch Science Center, Goethe Universit\"at Frankfurt, Ruth-Moufang-Str.\ 1, D-60438 Frankfurt am Main, Germany}

\author{Benjamin D\"onigus}
\affiliation{
Institut f\"ur Kernphysik,
Goethe Universit\"at Frankfurt, Max-von-Laue-Str.\ 1, D-60438 Frankfurt am Main, Germany}

\author{Behruz Kardan}
\affiliation{
Institut f\"ur Kernphysik,
Goethe Universit\"at Frankfurt, Max-von-Laue-Str.\ 1, D-60438 Frankfurt am Main, Germany}

\author{Manuel Lorenz}
\affiliation{
Institut f\"ur Kernphysik,
Goethe Universit\"at Frankfurt, Max-von-Laue-Str.\ 1, D-60438 Frankfurt am Main, Germany}

\author{Horst Stoecker}
\affiliation{
Institut f\"ur Theoretische Physik,
Goethe Universit\"at Frankfurt, Max-von-Laue-Str.\ 1, D-60438 Frankfurt am Main, Germany}
\affiliation{Frankfurt Institute for Advanced Studies, Giersch Science Center, Goethe Universit\"at Frankfurt, Ruth-Moufang-Str.\ 1, D-60438 Frankfurt am Main, Germany}
\affiliation{
GSI Helmholtzzentrum f\"ur Schwerionenforschung GmbH, Planckstr.\ 1, D-64291 Darmstadt, Germany}

\begin{abstract}
We estimate the feeddown contributions from decays of unstable $A = 4$ and $A = 5$ nuclei to the final yields of protons, deuterons, tritons, \tHe, and \fHe~produced in relativistic heavy-ion collisions at $\sqrt{s_{\rm NN}} > 2.4$~GeV, using the statistical model.
The feeddown contribution effects do not exceed 5\% at LHC and top RHIC energies due to the large penalty factors involved, but are substantial at intermediate collision energies.
We observe large feeddown contributions for tritons, \tHe, and \fHe~ at $\sqrt{s_{\rm NN}} \lesssim 10$~GeV, where they may account for as much as 70\% of the final yield at the lower end of the collision energies considered.
Sizable~($>10$\%) effects for deuteron yields are observed at $\sqrt{s_{\rm NN}} \lesssim 4$~GeV.
The results suggest that the excited nuclei feeddown cannot be neglected in the ongoing and future analysis of light nuclei production at intermediate collision energies, including HADES and CBM experiments at FAIR, NICA at JINR, RHIC beam energy scan and fixed-target programmes, and NA61/SHINE at CERN.
We further show that the freeze-out curve in the $T$-$\mu_B$ plane itself is affected significantly  by the light nuclei at high baryochemical potential.
\end{abstract}

\pacs{24.10.Pa, 25.75.Gz}

\keywords{statistical model, light nuclei production, excited nuclei}


\maketitle


\paragraph*{Introduction.}

The production of light nuclei, anti- and hypernuclei in relativistic heavy-ion collisions is an active topic of research centered around the studies of the QCD phase diagram.
The thermodynamic approach has been applied for a long time to describe fragment distributions in intermediate energy heavy-ion collisions ~\cite{Mekjian:1977ei,Mekjian:1978us,Mekjian:1978zz,Gosset:1988na,Siemens:1979dz,Stoecker:1984py,Hahn:1986mb,Csernai:1986qf,Jacak:1987zz,Hahn:1988zz}.
The basic model of particle production at relativistic energies is the hadron resonance gas~(HRG) model, which describes quite well the hadron yields measured at various energies ~\cite{Cleymans:1992zc,BraunMunzinger:1996mq,Becattini:2000jw}~(see, e.g., Ref.~\cite{Andronic:2017pug} for a recent overview). 
In the simplest case, the HRG model represents an ideal gas of non-interacting hadrons and resonances in the grand canonical ensemble.
One common extension of the HRG picture is to incorporate loosely-bound objects such as light (anti-)(hyper-)nuclei. Within the ideal HRG model these objects are implemented as point-like, non-interacting particles carrying their quantum numbers and masses. The model thus provides essentially a parameter-free prediction of light nuclei yields in relativistic heavy-ion collisions which, in most cases, is in a remarkably good agreement with experimental data~\cite{BraunMunzinger:1994iq,Andronic:2010qu,Steinheimer:2012tb,Adam:2015vda,Adam:2015yta,Anticic:2016ckv}.

In addition to the well known light nuclei that are stable under strong interactions~(d, \trit, \tHe, \fHe, etc.) also a large number of excited nuclear states is established~\cite{NNDC,NSR2012WA38}.
These states start from the mass number $A = 4$.
The importance of excited nuclear states and their feeddown is, of course, well known for the description of light nuclei formation in intermediate energy nuclear collisions in the spectator region~\cite{Fai:1982zk,Fai:1983pd,Hahn:1986mb,Bondorf:1995ua}.
This feeddown, however, is seldom considered in heavy-ion reactions at ultrarelativistic energies, such as those studied at SPS, RHIC and LHC.
A strong suppression of yields of heavy particles makes the feeddown from their decays to lighter particles considerably suppressed.
The penalty factor, i.e. the suppression of the yield of a nucleus that results by adding one nucleon to its content, in the thermal model is given by $P \sim \exp[(m_N - \mu_B)/T]$, where $T$ and $\mu_B$ are the temperature and baryochemical potential at the freeze-out and $m_N$ is the nucleon mass. 
The penalty factor is larger than 10 at $\mu_B < 600$~MeV and reaches the values of order 300-400 at the LHC.

On the other hand, as recently been pointed out~\cite{Shuryak:2019ikv}, the number of excited $^4$He nuclei is quite large: about 50 different spin and excitation energy states are known~\cite{NNDC,NSR2012WA38}.
In the statistical model all these states are populated almost equally, given the smallness of their level spacing~($\sim 10$~MeV) to the typical chemical freeze-out temperatures $T \sim 100-150$~MeV.
All these excited states feed into the stable lower mass nuclei: \tHe, \trit, and d.
The large number of excited states can thus compensate the large penalty factor.
In Ref.~\cite{Shuryak:2019ikv} estimates were presented of how much nucleons, deuterons, \tHe, and \trit~are produced on average from a decay of an excited \fHe~state. 
Here we extend these considerations into a full thermal model calculation, incorporating, in addition to \fHe~states, also $^4$H, $^4$Li as well as $^5$H, $^5$He and $^5$Li excited states.
We present a quantitative estimate of the feeddown contributions along the chemical freeze-out curve in heavy-ion collisions.

\paragraph*{Excited light nuclei in HRG model.}

We use the ideal hadron resonance gas~(HRG) model to describe the various particle abundances at the chemical freeze-out~\cite{Cleymans:1992zc,BraunMunzinger:1996mq,Becattini:2000jw}.
The final multiplicity of particle species $i$ consists of the primordial yield and feeddown contributions:
\eq{
N_i^{\rm tot} = N_i^{\rm prim} + N_i^{\rm feed}.
}
Here
\eq{
N_i^{\rm prim} = V \, \frac{d_i \, m_i^2 \, T^2}{2\pi^2} \, K_2\left(\frac{m_i}{T}\right) \, \exp\left(\frac{\mu_i}{T}\right),
}
where $d_i$ and $m_i$ are particle's spin degeneracy and mass, respectively. 
$\mu_i$ is the chemical potential of the particle species $i$ which is determined by the baryon number, electric charge and strangeness chemical potentials, $\mu_i = \mu_B b_i + \mu_Q q_i + \mu_S s_i$.
The feeddown contributions include all strong and electromagnetic decays of resonances and excited nuclei:
\eq{
N_i^{\rm feed} = \sum_j \, \langle n_i \rangle_j \, N_j^{\rm prim}.
}
Here $\langle n_i \rangle_j$ is the average number of particles
of type $i$ resulting from decay of unstable particle of type $j$ and subsequent decay chains of all unstable decay products.

We use an open source thermal model package \texttt{Thermal-FIST}~\cite{Vovchenko:2019pjl} in our analysis.
We expand the default PDG2014 particle list in \texttt{Thermal-FIST} by adding $A = 4$ and $A = 5$ excited nuclear states and their branching ratios from Refs.~\cite{Tilley:1992zz} and~\cite{Tilley:2002vg}, respectively.
We incorporate only the states with well established spin and parity $J^\pi$, energy level, and decay branching ratios.
The list of all such states is shown in tables~\ref{tab:A4} and~\ref{tab:A5}.
The $^4$H and $^4$Li sectors contain the same number of $T = 1$ states, related to each other by isospin symmetry, while the $^4$He sector contains both the $T = 0$ and $T = 1$ channels, yielding a considerably larger amount of excited nuclear states.

\begin{table}
 \caption{Level chart of $A = 4$ nuclear states. The ground state (g.s.) energy for $^4$He corresponds to the $\alpha$ particle mass. The g.s. energies of $^4$H and $^4$Li states lie 3.19 MeV above $\text{n} + \text{t}$ and 4.07 MeV above $\text{p}+^3\text{He}$ threshold masses, respectively. All excited states decay into stable lighter nuclei and nucleons, the last column specifies the branching ratios corresponding to emission of various particles. } 
 \centering                                                 
 \begin{tabular}{|c|c|c|c|}   
 \hline
 $A = 4$ & $E_x$(MeV) & $J^\pi$ & Decay channels \\
 \hline
 \hline
 $^4$H & g.s. & $2^-$ & n(100\%) \\
       & 0.31 & $1^-$ & n(100\%) \\
       & 2.08 & $0^-$ & n(100\%) \\
       & 2.83 & $1^-$ & n(100\%) \\
 \hline
 \hline
 $^4$He & g.s.  & $0^+$ & stable \\
        & 20.21 & $0^+$ & p(100\%) \\
        & 21.01 & $0^-$ & n(23.8\%), p(76.2\%) \\
        & 21.84 & $2^-$ & n(37.3\%), p(62.7\%) \\
        & 23.33 & $2^-$ & n(47.3\%), p(52.7\%) \\
        & 23.64 & $1^-$ & n(44.5\%), p(55.5\%) \\
        & 24.25 & $1^-$ & n(47.0\%), p(50.5\%), d(2.5\%) \\
        & 25.28 & $0^-$ & n(48.3\%), p(51.7\%) \\
        & 25.95 & $1^-$ & n(48.5\%), p(51.5\%) \\
        & 27.42 & $2^+$ & n(3\%), p(3\%), d(94\%) \\
        & 28.31 & $1^+$ & n(47\%), p(48\%), d(5\%) \\
        & 28.37 & $1^-$ & n(2\%), p(2\%), d(96\%) \\
        & 28.39 & $2^-$ & n(0.25\%), p(0.25\%), d(99.5\%) \\
        & 28.64 & $0^-$ & d(100\%) \\
        & 28.67 & $2^+$ & d(100\%) \\
        & 29.89 & $2^+$ & n(0.4\%), p(0.4\%), d(99.2\%) \\
 \hline
 \hline
 $^4$Li & g.s. & $2^-$ & p(100\%) \\
        & 0.32 & $1^-$ & p(100\%) \\
        & 2.08 & $0^-$ & p(100\%) \\
        & 2.85 & $1^-$ & p(100\%) \\
 \hline
 \end{tabular}
\label{tab:A4}
\end{table}

\begin{table}
 \caption{Level chart of $A = 5$ nuclear states. The ground state (g.s.) energy for $^5$H corresponds to 4.69104 GeV. 
 The g.s. energies for $^5$He and $^5$Li lie 0.798 MeV above $\text{n} + \alpha$ and 1.69 MeV above $\text{p}+\alpha$ threshold masses, respectively.} 
 \centering                                                 
 \begin{tabular}{|c|c|c|c|}   
 \hline
 $A = 5$ & $E_x$(MeV) & $J^\pi$ & Decay channels \\
 \hline
 \hline
 $^5$H  & g.s.  & ${1 \over 2}^+$ & 2n(100\%) \\
 \hline
 \hline
 $^5$He & g.s.  & ${3 \over 2}^-$ & n(100\%) \\
        & 1.27  & ${1 \over 2}^-$ & n(100\%) \\
        & 16.84 & ${3 \over 2}^+$ & n(60\%), d(40\%) \\
 \hline
 \hline
 $^5$Li & g.s.  & ${3 \over 2}^-$ & p(100\%) \\
        & 1.49  & ${1 \over 2}^-$ & p(100\%) \\
        & 16.87 & ${3 \over 2}^+$ & p(70\%), n(30\%) \\
 \hline
 \end{tabular}
\label{tab:A5}
\end{table}

The excited nuclei decay into stable lighter nuclei and nucleons, with typical decay widths of order $\Gamma = 1-10$~MeV.
$^4$H and $^4$Li are assumed to decay into t$ + $n and $^3$He$+$p, respectively, with 100\% probability.
$^4$He states decay into t$ + $p, d$+$d, $^3$He$+$n, with different branching ratios in the right column of Table~\ref{tab:A4}.
For the $A = 5$ states the channels $^5\text{H} \to~ \text{t} + \text{n} + \text{n}$, $^5\text{He} \to~ ^4\text{H} + \text{p}$, $^5\text{He} \to~ \text{t} + \text{d}$, $^5\text{Li} \to~ ^4\text{He} + \text{p}$, and $^5\text{Li} \to~ ^4\text{Li} + \text{n}$ are taken into account.
The excited nuclei feeding will thus affect the yields of nucleons and stable light nuclei, but not other particles.

Each collision energy is characterized by three freeze-out parameters: the temperature $T_{\rm ch}$, the baryochemical potential $\mu_B^{\rm ch}$, and the volume $V_{\rm ch}$.
The volume parameter cancels out in all intensive quantities, such as a ratio of any two yields.
The temperature and the baryochemical potentials can be mapped into the collision energies phenomenologically, through a thermal model analysis of hadron yields at various collision energies~\cite{Cleymans:2005xv,Andronic:2005yp}.
In the present study we shall use the chemical freeze-out curve of Ref.~\cite{Vovchenko:2015idt}.
The electric charge and strangeness chemical potentials $\mu_Q$ and $\mu_S$ are determined at each energy to satisfy the conservation laws given by the content of the incoming nuclei: the electric to baryon charge ratio, $Q/B = 0.4$, and the vanishing net strangeness, $S = 0$.

\begin{figure}[t]
  \centering
  \includegraphics[width=.47\textwidth]{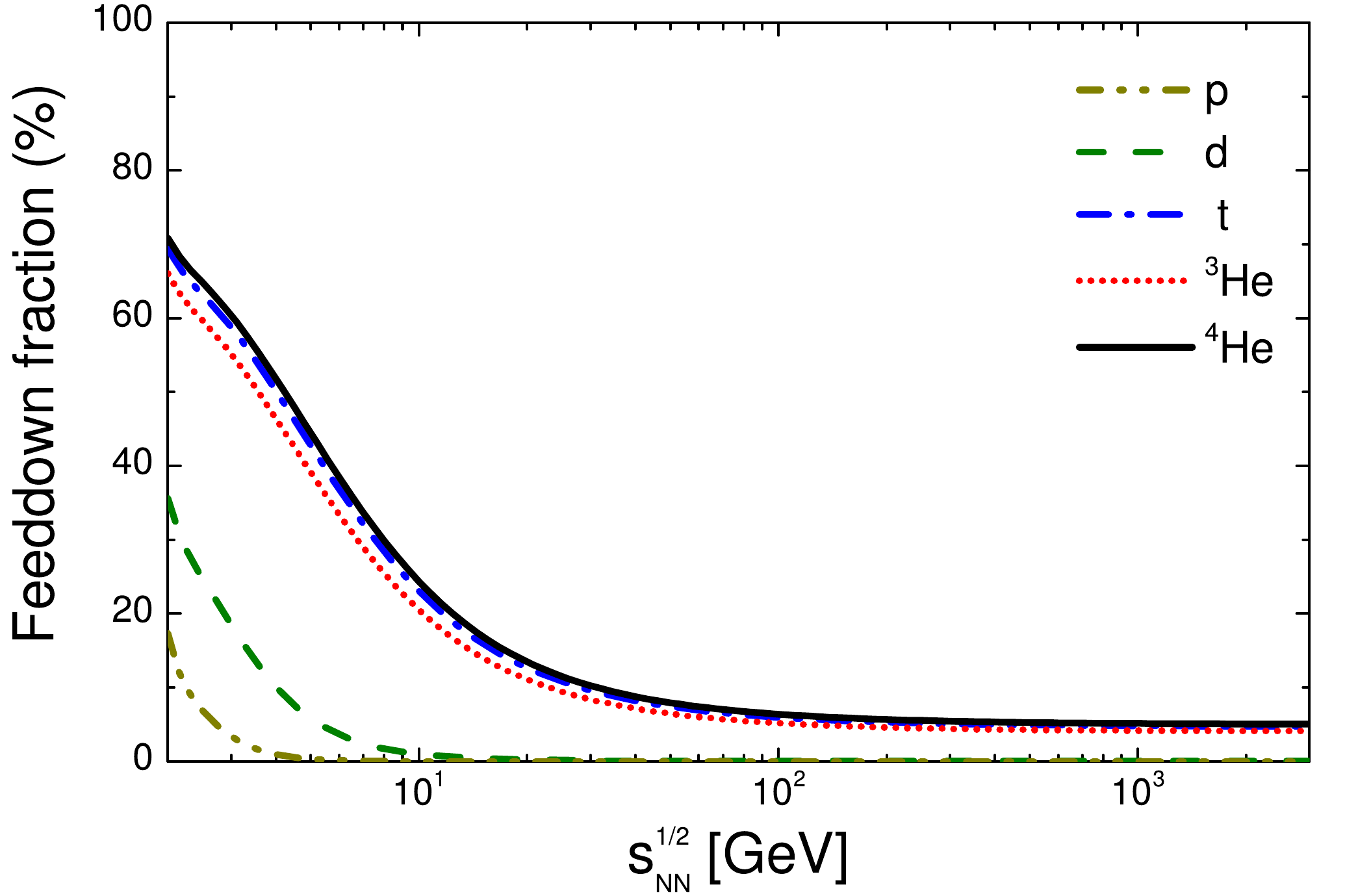}
  \caption{\label{fig:feeddown}
  Fractions of final yields of p, d, t, $^3$He, and $^4$He coming from decays of the excited nuclei estimated in the statistical model along the chemical freeze-out curve of Ref.~\cite{Vovchenko:2015idt}.
  }
\end{figure}

Figure~\ref{fig:feeddown} depicts the collision energy dependence of the fraction of final yields for p, d, t, $^3$He, $^4$He coming from decays of excited nuclear states, 
given as the ratio $N_i^{\rm feed} / N_i^{\rm tot}$.
The feeddown fraction is a monotonically decreasing function of the collision energies, saturating at high energies.
This is mainly correlated with the monotonic decrease of the baryochemical potential $\mu_B$ which approaches $\mu_B \simeq 0$ at the highest collision energies available at LHC. \\
At the LHC energies the feeddown from individual excited states is strongly suppressed due to the large penalty factor. 
The feeddown to deuteron yields comes mainly from decays of a number of excited $^4$He states~(see Table~\ref{tab:A4}).
As the nucleon mass number difference between d and $^4$He is two, this feeddown is suppressed by the square of the penalty factor, $N_{\text{d}}^{\rm feed} \sim P^{-2}$.
The feeddown is thus negligible at high energies, and only starts to contribute to the deuteron yield at $\sqrt{s_{\rm NN}} \lesssim 5$~GeV.
The standard thermal model is thus sufficient for the analysis of d abundances at SPS, RHIC, and LHC, but feeddown corrections are necessary at lower energies, e.g. at HADES \cite{Agakishiev:2009am}.\\
The main contributions to t and $^3$He come from decays of $^4$H, $^4$He, and $^4$Li states whereas the decays of $^5$He and $^5$Li feed substantially into the yields of $^4$He.
Due to a large number of excited states~(see Tables~\ref{tab:A4} and~\ref{tab:A5}), a 5\% effect 
to the yields of t, $^3$He, and $^4$He is observed in our calculations at the highest collision energies.
The future high-luminosity measurements at Runs 3 and 4 at the LHC~\cite{Citron:2018lsq} are potentially sensitive to these effects. 
Prospective future measurements of excited nuclei abundances at the LHC can serve as an additional test of thermal production mechanism for loosely-bound objects in high energy collisions.

At energies, $\sqrt{s_{\rm NN}} \lesssim 10$~GeV, the feeddown contributions to the yields of t, $^3$He, and $^4$He are substantial and reach up to 70\% of the final yield, while the one to the d reaches up to 30\%.
These results are consistent with the quantum statistical model of fragment formation analysis in Ref.~\cite{Hahn:1986mb}, where intermediate collisions energies with entropy per baryon ratio values $S/A \lesssim 5$~($\sqrt{s_{\rm NN}} \lesssim 2.4$~GeV) were considered.
It is therefore important to incorporate the excited states for a quantitative analysis of light nuclei production in such heavy-ion experiments as beam energy scan programmes at RHIC and SPS, the HADES experiment at GSI, as well the future experiments CBM at FAIR and BM@N at NICA.

\begin{figure}[t]
  \centering
  \includegraphics[width=.47\textwidth]{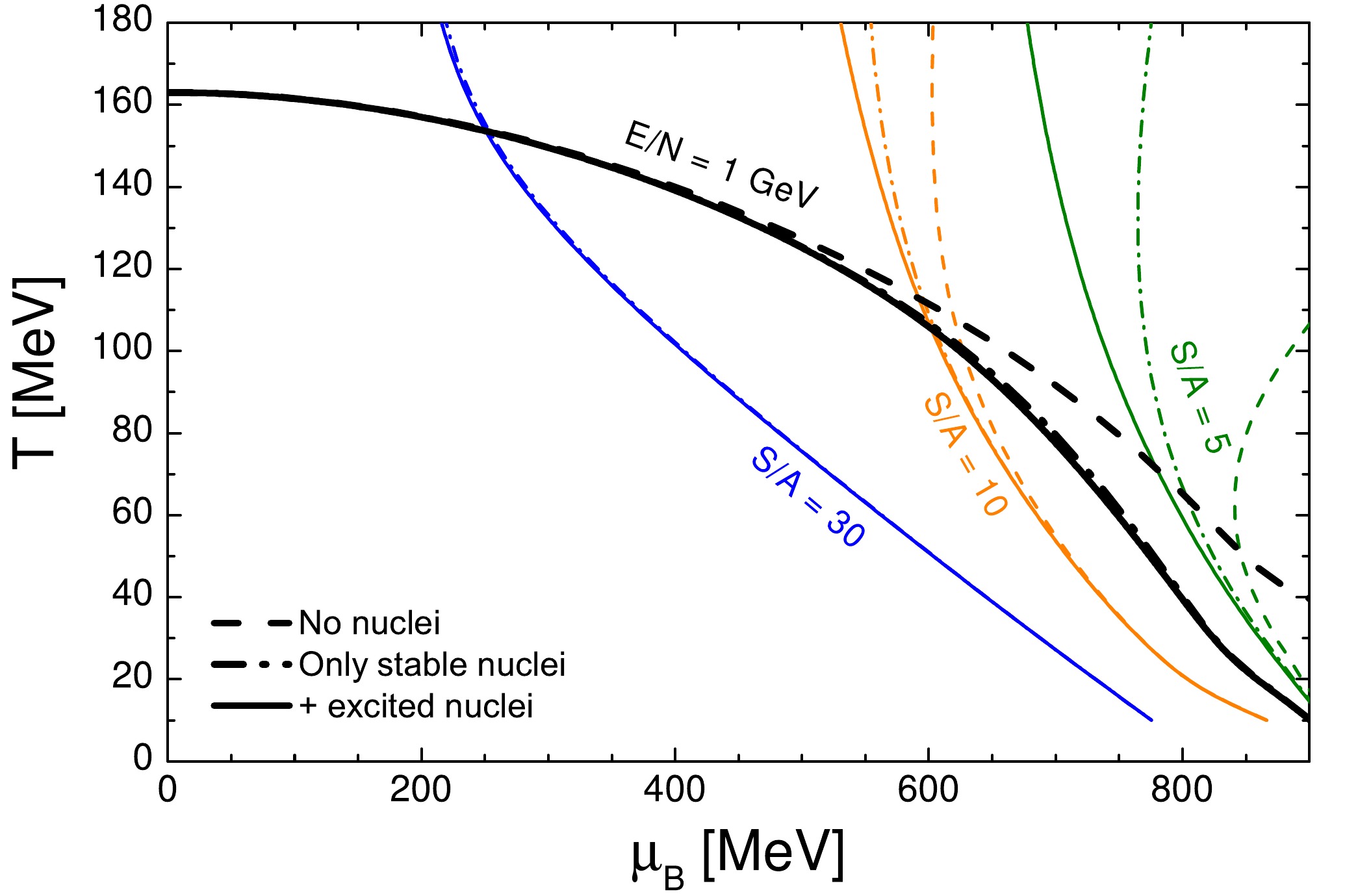}
  \caption{\label{fig:Tmu}
  Black lines: constant energy per particle ratio $E/N = 1$~GeV for the case of no nuclei~(dashed), ground state nuclei~(dash-dotted) and excited nuclei~(solid) being included in the statistical model calculation.
  The colored lines correspond to constant entropy per baryon trajectories: $S/A = 5$~(green),  $S/A = 10$~(orange), and $S/A = 30$~(blue).
  }
\end{figure}

Indeed, for those  energies  the  freeze-out curve in the $T$-$\mu_B$ plane can to large extent be defined by the light nuclei themselves, where they constitute a large fraction of measured particles.
Figure~\ref{fig:Tmu} visualizes the impact of the inclusion of light nuclei on the energy per particle $E/N=1$ GeV curve, which is one of the proposed universal freeze-out criteria for heavy-ion collisions~\cite{Cleymans:1998fq,Cleymans:1999st}\footnote{In fact, Refs.~\cite{Cleymans:1998fq,Cleymans:1999st} used the data involving light nuclei, namely the d/p ratio at low energies, to obtain this criterion.}. 
Again, at low baryochemical potential the effect is negligible while from $\mu_{B} \simeq 500$~MeV on, differences between the curves with and without nuclei become visible. At a given $\mu_{B} \simeq 800$~MeV the difference in temperature $T$ is almost a factor 2 between the version without nuclei and the ones including nuclei.
Also shown in Fig.~\ref{fig:Tmu} are constant entropy per baryon trajectories for different values of $S/A$.
The effect of including the stable and excited light nuclei is most visible in the baryon-rich region, e.g. for $S/A = 5$ which is an approximate phase diagram trajectory of the HADES experiment.

We point out that in the present version of \texttt{Thermal-FIST} excited states are included only until $A=5$. If the baryochemical potential $\mu_{B}$ approaches the nucleon mass $\mu_{B} \simeq M_{N}$, the population of heavier nuclei starts to explode and the effects of the inclusion of excited nuclei states in addition, might also become visible in the $E/N=1$~GeV curve. 
Furthermore, dynamics associated with the presence of the nuclear liquid-gas transition~\cite{Pochodzalla:1995xy} become increasingly important in cold and dense region of the phase diagram, and the HRG model would have to be extended to incorporate this~\cite{Vovchenko:2016rkn,Poberezhnyuk:2019pxs}.
We will study these effects in a greater detail in a further publication.

The decays of excited nuclei also play role in the total abundances of nucleons, although the effect is notable only at low collision energies, $\sqrt{s_{\rm NN}} \lesssim 3$~GeV.
For instance, about 10\% of final proton yield comes from decays of excited nuclei at a HADES energy of $\sqrt{s_{\rm NN}} = 2.4$~GeV.

\begin{figure}[t]
  \centering
  \includegraphics[width=.47\textwidth]{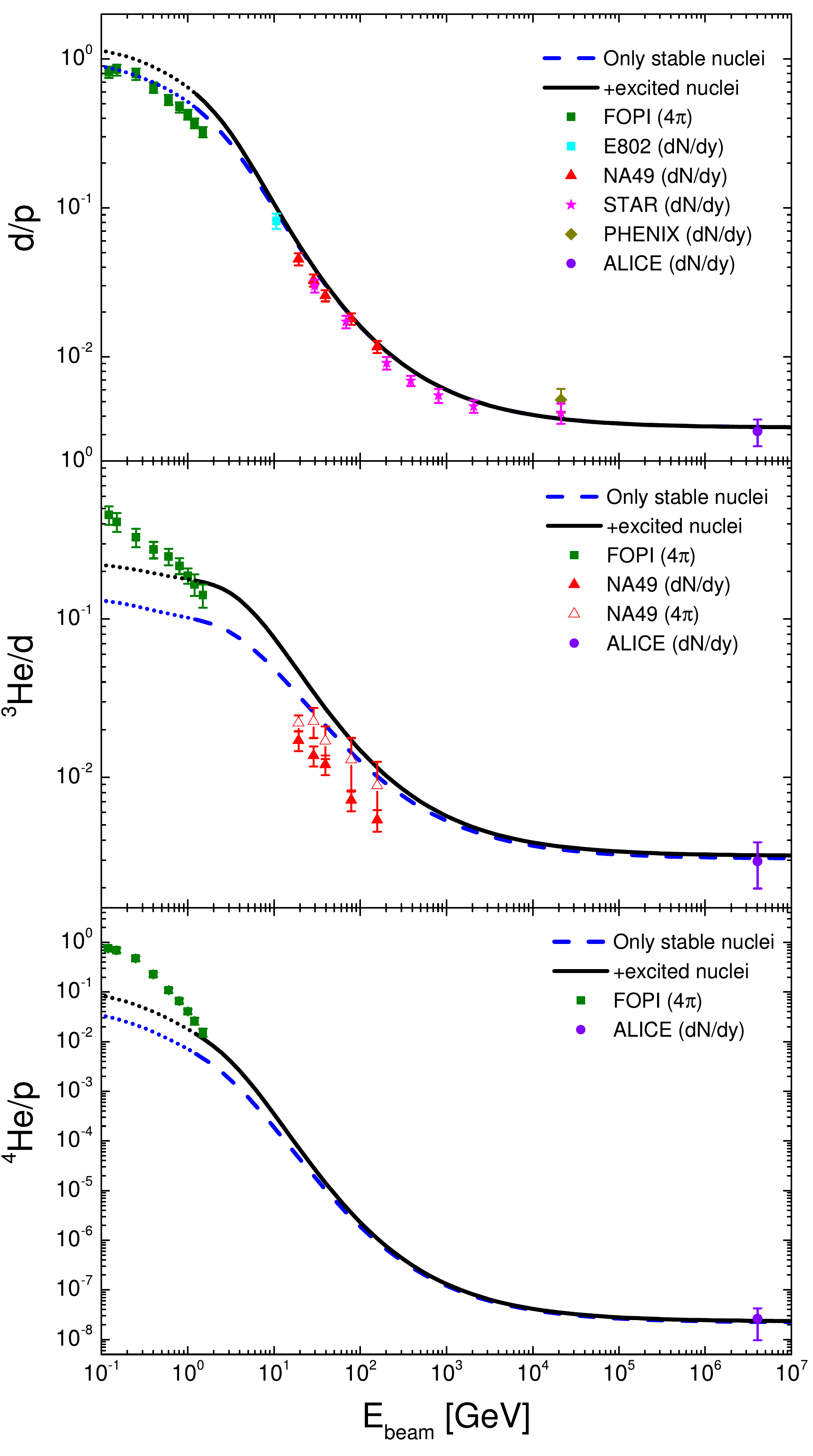}
  \caption{\label{fig:ratios}
  Beam energy dependence of the ratios d/p, \tHe/d, \fHe/p~(from top to bottom), estimated in the statistical model along the chemical freeze-out curve of Ref.~\cite{Vovchenko:2015idt} compared to the world data.
  The black solid lines correspond to calculations including both the ground state and excited nuclei whereas the dashed blue lines depict calculations that include ground state nuclei only. The dotted lines correspond to extrapolating the freeze-out curve to low collision energies, $\sqrt{s_{\rm NN}} < 2.4$~GeV.
  }
\end{figure}

\paragraph*{Yield ratios.}

For model-to-data comparison we focus on the d/p, \tHe/d, and \fHe/p yield ratios, for which rich data sets exist. 
Figure~\ref{fig:ratios} depicts the collision energy dependence of the yield ratios with and without the inclusion of excited nuclear states. As can be seen, when comparing the dashed blue and the solid black lines, in terms of feeddown the \tHe/d and \fHe/p ratios are more interesting where the effect is a considerable one.
On the other hand, more data are available for the d/p ratio. \\
The ALICE data~\cite{Adam:2015vda} for the d/p, \tHe/d, and \fHe/p ratios are described fairly well by the thermal model. 
While at 200 GeV the model is in favor of the STAR~\cite{Adam:2019wnb} over the PHENIX~\cite{Adler:2004uy,Adler:2003cb} d/p data, at intermediate energies there is a trend to slightly overshoot both the NA49~\cite{Anticic:2016ckv} and STAR data~\cite{Adam:2019wnb}. Looking at the \tHe/d ratio at intermediate energies, one sees that the thermal model, with or without nuclear feeddown, overestimates significantly the NA49 data for the \tHe/d ratio of yields measured at midrapidity.
A similar observation has also been recently made for the preliminary beam energy scan STAR data for the d$N$/d$y$ yield ratio \trit/d~\cite{STARprelim}.
It is evident from the NA49 data itself, that the \tHe/d ratio constructed from $4\pi$ yields is considerably higher than the one measured at midrapidity.
This difference comes from an increase of rapidity densities of light nuclei as one goes away from the midrapidity slice~\cite{Anticic:2016ckv}. 
The effect is larger for $^3$He than for deuterons.
In the framework of thermal model, the rapidity dependence of yield ratios can be understood in terms of rapidity dependence of the chemical freeze-out temperature and baryochemical potential and the presence of the longitudinal flow.
More specifically, following Refs.~\cite{Biedron:2006vf,Broniowski:2007mu,Becattini:2007ci} one assumes that the longitudinal rapidity axis at freeze-out is populated by various fireballs, the temperature and baryochemical potentials for each fireball lie on the chemical freeze-out curve but the exact location depends on both the collision energy and the fireball rapidity $Y_{\rm FB}$.
To leading order, the dependence of the baryochemical potential on $Y_{\rm FB}$ is quadratic: $\mu_B(Y_{\rm FB}) = \mu_B(0) + b \, Y_{\rm FB}^2$.
Positive values $b > 0$ are suggested by the analysis of (net-)proton rapidity distributions at $\sqrt{s_{\rm NN}} = 17.3$ and 200~GeV~\cite{Becattini:2007ci}, implying stronger increase of rapidity densities of particles with a higher baryon content as one moves away from midrapidity.

\begin{figure}[t]
  \centering
  \includegraphics[width=.47\textwidth]{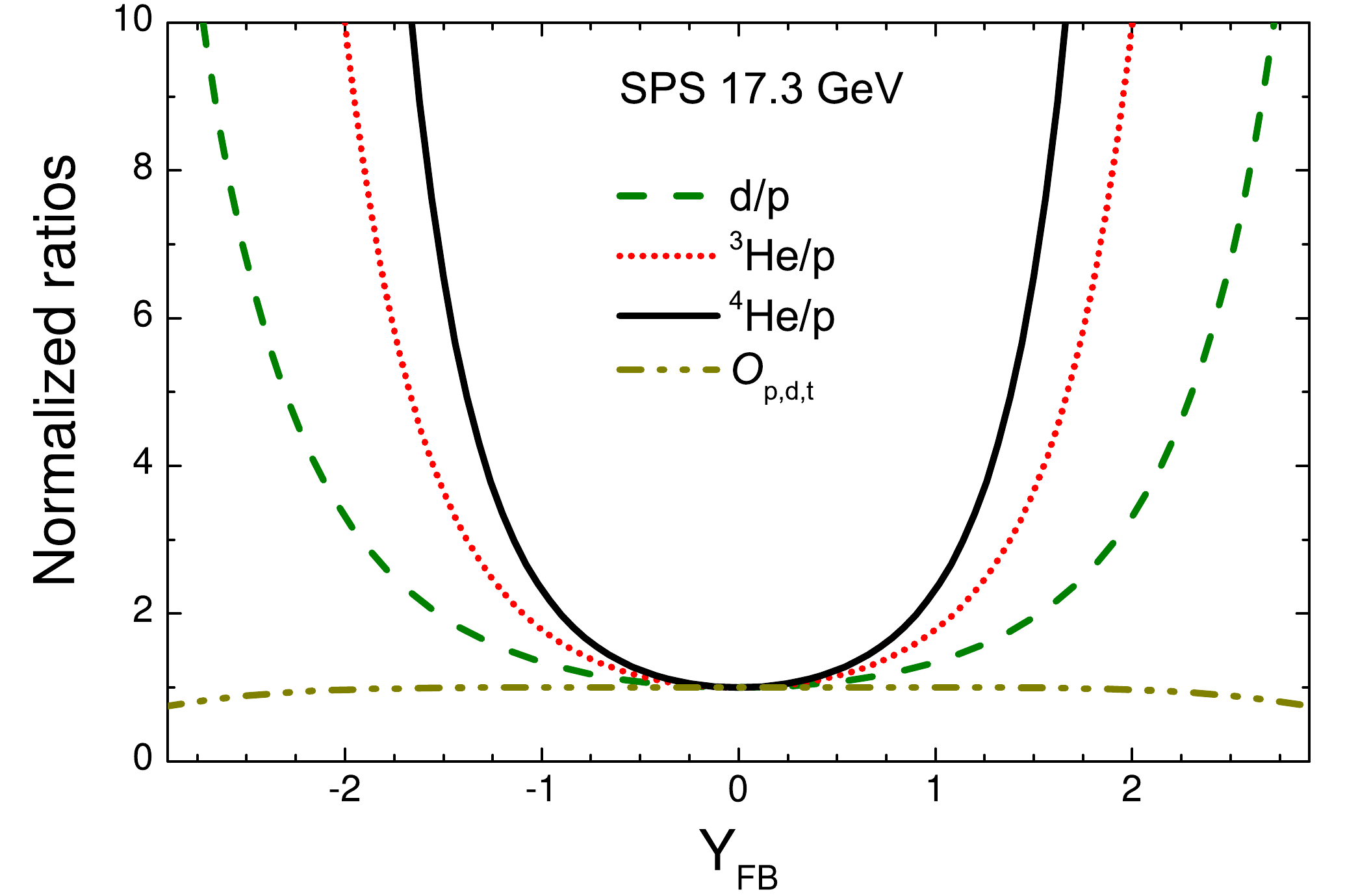}
  \caption{\label{fig:rapidity}
  Dependence of d/p, \tHe/p, \fHe/p, and $O_{\text{p},\text{d},\text{t}}$ ratios on the fireball rapidity $Y_{\rm FB}$ at top SPS energy of 17.3~GeV assuming parabolic $Y_{\rm FB}$-dependence of baryochemical potential with parameters taken from~Ref.~\cite{Becattini:2007ci}.
  The ratios are normalized by their values at $Y_{\rm FB} = 0$.
  }
\end{figure}

To illustrate this effect we show in Fig.~\ref{fig:rapidity} the $Y_{\rm FB}$-dependence of the ratios d/p, \tHe/p, \fHe/p, 
calculated within \texttt{Thermal-FIST} for $\sqrt{s_{\rm NN}} = 17.3$~GeV using parameters $\mu_B(0) = 237$~MeV and $b = 50$~MeV from Ref.~\cite{Becattini:2007ci}.
The light nuclei abundances show a rapid increase with $Y_{\rm FB}$ relative to protons, with a more pronounced effect for heavier nuclei. 
The ratios stay relatively flat only in a narrow region around midrapidity that does not exceed 1-2 units.
Therefore, the mapping of chemical freeze-out parameters to collision energy depends on whether midrapidity or $4\pi$ data are considered.
Differences between the two are more pronounced for light nuclei.
We note that the chemical freeze-out curve of Ref.~\cite{Vovchenko:2015idt} was obtained using the $4\pi$ hadron yields of the NA49 collaboration and, therefore, it is more suited to analyze the $4\pi$ rather than d$N$/d$y$ yields in that energy range.
Whereas, the d/p data point from the E802 collaboration~\cite{Ahle:1999in} is compatible with the curve and originates from d$N$/d$y$ values. It is worth to mention that, there is nice data from the same energy as E802 form the AGS, which is not easy to compare to our predictions since it focuses on extrapolations towards transverse momenta of zero and not give any integrated d$N$/d$y$ or $4\pi$ yields (see Refs.~\cite{Barrette:1999kq,Armstrong:2000gz,Armstrong:2000gd}). These data are nevertheless described by the thermal model approach~\cite{BraunMunzinger:2001mh}.

The inclusion of the feeddown, mainly from the decays of $A = 4$ nuclei, increases the \tHe/d ratio and worsens the agreement with the data for the lowest three NA49 energies. \\
At low energies, where $4\pi$ data from the FOPI collaboration are available~\cite{Reisdorf:2010aa} and the effect of the feeddown is the largest, no clear picture emerges. While, the d/p data are in better agreement with the model without the inclusion of the excited states, the data for both the \tHe/d and \fHe/p ratios clearly favor the inclusion of these states. The reason for this can be manifold. As already pointed out, at those energies the freeze-out is to large extend defined by the light nuclei and their treatment. 
Inclusion of larger and heavier excited states than those considered in the present work is likely necessary.
The situation is additionally complicated by the need for a canonical calculation of strangeness \cite{Cleymans:1998yb} and a general lack in the variety of measured hadron yields.
The application of the freeze-out curve in the FOPI energy range is an extrapolation from higher energies and cannot be considered to be fully controlled. 
No data below $\sqrt{s_{\rm NN}} \simeq 2.4$~GeV were used in Ref.~\cite{Vovchenko:2015idt} to construct the freeze-out curve, for this reason we depict in Fig.~\ref{fig:ratios} the results at these energies by dotted lines.
In addition to a more reliable freeze-out concept at low energies, the  present model will need to be supplemented by additional excited states~\cite{Jacak:1983iz,Hahn:1986mb,Jacak:1987zz}, as well as the effects due to nuclear mean fields. This will be in the focus of a forthcoming publication.
Another problem is a difficulty to estimate the contribution of multifragmentation in the $4\pi$ data, since it includes a given fraction in the forward region. This becomes even more important at energies below SIS energies.  
New high-quality data is thus eagerly awaited \cite{MLo,Mszala}. 
In addition, we point the interested reader to various measurements of particle unstable nuclei through correlation functions \cite{Pochodzalla:1985zz,Pochodzalla:1986dqq,Kotte:1999gr,Finch:1999ja} at low energies, which might be used shed additional light on the situation.

A particularly interesting observable with regards to the light nuclei production is a (double) ratio $O_{\text{p},\text{d},\text{t}} = N_\text{p} \, N_{\text{t}} / N_\text{d}^2$, proposed recently in Refs.~\cite{Sun:2017xrx,Sun:2018jhg}.
The dependence on the chemical potentials $\mu_B$, $\mu_Q$, and $\mu_S$ drops out in $O_{\text{p},\text{d},\text{t}}$ to a leading order, as follows from the $B,Q,S$ contents of $\text{p}$, $\text{d}$, and t. This results in an almost flat expected rapidity dependence of this ratio~(Fig.~\ref{fig:rapidity}).
Furthermore, using the coalescence model it was shown that $O_{\text{p},\text{d},\text{t}}$ can be related to neutron density fluctuations and thus be sensitive to the hypothetical QCD critical point, which would be signalled by a non-monotonous collision energy dependence of $O_{\text{p},\text{d},\text{t}}$.
Recent work~\cite{Shuryak:2019ikv} has indicated that $O_{\text{p},\text{d},\text{t}}$ is potentially sensitive to the feeddown from decays of excited \fHe~states.

\begin{figure}[t]
  \centering
  \includegraphics[width=.47\textwidth]{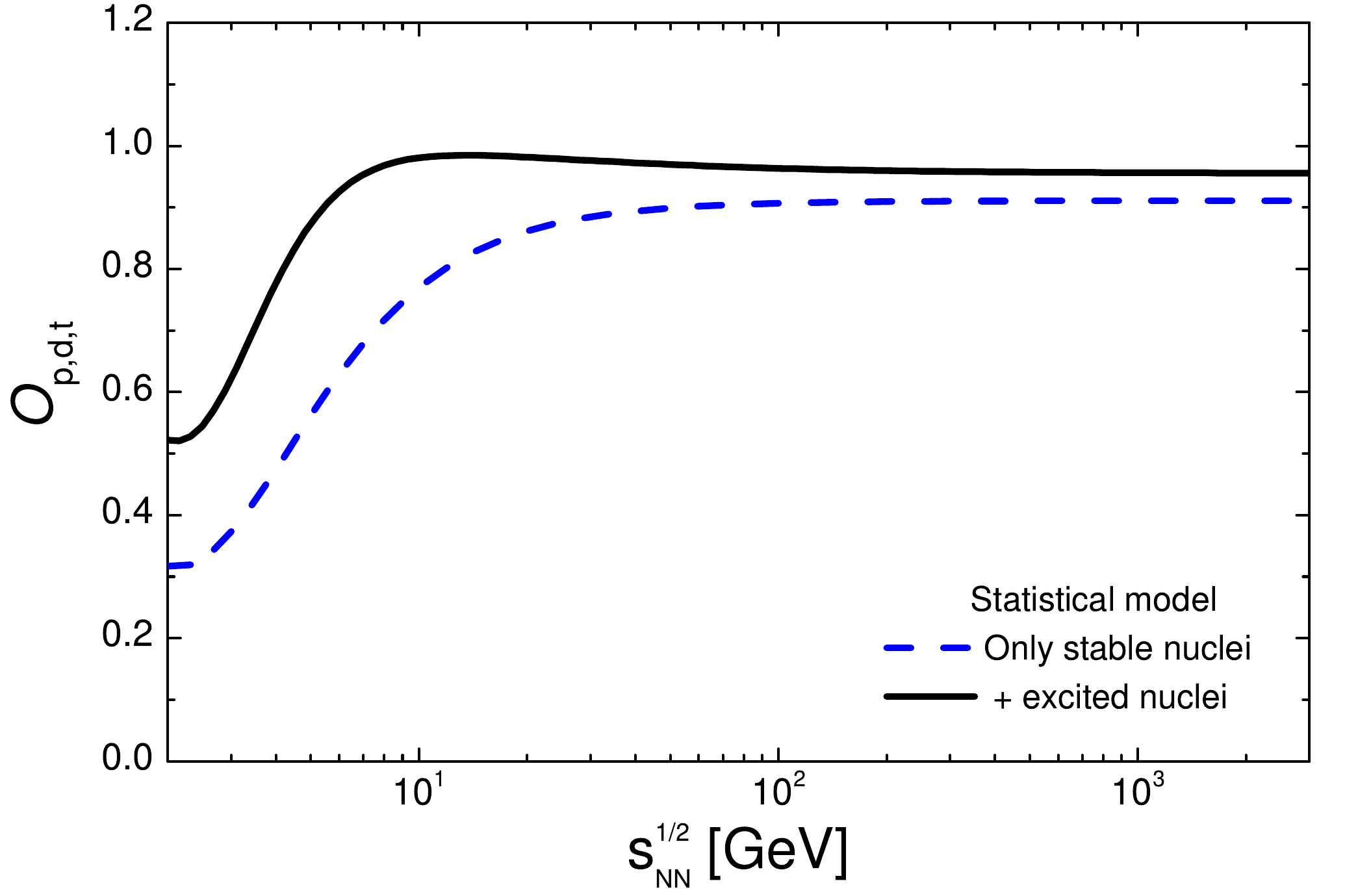}
  \caption{\label{fig:Opdt}
  Collision energy dependence of the ratio $O_{\text{p},\text{d},\text{t}} = N_\text{p} \, N_\text{t} / N_\text{d}^2$ estimated in the statistical model along the chemical freeze-out curve of Ref.~\cite{Vovchenko:2015idt}.
  }
\end{figure}

The collision energy dependence of $O_{\text{p},\text{d},\text{t}}$ evaluated in the thermal model is shown in Fig.~\ref{fig:Opdt}, with~(solid black line) and without~(dashed blue line) feeddown from excited nuclei decays.
If no excited nuclei are included, $O_{\text{p},\text{d},\text{t}}$ exhibits a monotonic increase with collision energy which saturates at $\sqrt{s_{\rm NN}} \gtrsim 20$~GeV at $O_{\text{p},\text{d},\text{t}} \simeq 0.9$.
The main reason for the collision energy dependence of $O_{\text{p},\text{d},\text{t}}$ is the increasing role of baryonic resonances~(and their feeddown to protons) as $\sqrt{s_{\rm NN}}$ is increased.

A stronger collision energy is observed when decays of the excited $A = 4$ states are incorporated~(solid black line in Fig.~\ref{fig:Opdt}). 
$O_{\text{p},\text{d},\text{t}}$ in this case is considerably larger at intermediate collision energies, $\sqrt{s_{\rm NN}} \lesssim 20$, where feeddown to \text{t} is significant.
At high energies, $O_{\text{p},\text{d},\text{t}}$ saturates at a value of about 0.96.
Due to the feeddown from excited nuclear states, $O_{\text{p},\text{d},\text{t}}$ exhibits a non-monotonic  $\sqrt{s_{\rm NN}}$ dependence, with a broad peak centered around $\sqrt{s_{\rm NN}} \simeq 10$~GeV.
Even though the obtained peak is not pronounced, this result demonstrates that a non-monotonic collision energy dependence of $O_{\text{p},\text{d},\text{t}}$ is possible without incorporating effects associated with the QCD critical point.

We do not present a comparison of our results for $O_{\text{p},\text{d},\text{t}}$ with experimental data.
The data for \trit~yields in all the relevant experiments are either still preliminary~(HADES, STAR, ALICE), or has to be extrapolated from the measured \tHe~yields~(NA49).
The current status of preliminary data on $O_{\text{p},\text{d},\text{t}}$ and their comparison with different theoretical predictions has recently been presented in Ref.~\cite{Oliinychenko:2020ply}, showing that all presently available models struggle to describe the data.
The results of our paper indicate that the excited nuclei shall play an important role in the on-going efforts to understand the behavior of this ratio at various collision energies.

Before we conclude, we would like to discuss the conceptual issues surrounding the thermal model approach to light nuclei production, recent advances in that direction, and how this affects the excited nuclear states.
Given that the nuclear binding energies are of order of few MeV or less, they are much smaller than the thermal energies encountered at the chemical freeze-out stage of heavy-ion collisions, indicating that the loosely-bound states are not supposed to be able to exist at that stage, or, if they do, to survive all the way to their detection. A recent discussion of these issues can be found in Ref.~\cite{Cai:2019jtk}.
Nevertheless, the thermal model provides a remarkable description of the experimentally measured nuclear abundances.
A promising avenue toward the resolution of this puzzle was recently put forward in Refs.~\cite{Oliinychenko:2018ugs,Xu:2018jff,Vovchenko:2019aoz}: 
the light nuclei abundances are not frozen at the conventional chemical freeze-out, but undergo destruction and regeneration reactions in the hadronic phase.
The large inelastic pion-nucleus reaction cross sections maintain the detailed balance throughout the hadronic phase, resulting in final nuclear abundances that are within 10-20\% of the thermal model values.
This observation explains the success of the thermal model in describing the data, although it does not justify the assumptions of the model.

To establish the possible influence of the hadronic phase on excited nuclei feeddown we perform a calculation of excited nuclei abundances in the framework of partial chemical equilibrium~\cite{Bebie:1991ij} at LHC energies.
The calculation setup is the same as in Ref.~\cite{Vovchenko:2019aoz}, the only difference being the addition of excited nuclei to the particle list.
In the spirit of partial chemical equilibrium, we express the excited nuclei fugacities at a given temperature in the hadronic phase in terms of the fugacities of their decay products~\cite{Bebie:1991ij}.
We verified that all abundances of excited nuclei stay within 10\% of the chemical freeze-out value at any reasonable temperature characterizing the hadronic phase: $90 < T < 155$~MeV.
Therefore, we expect our excited nuclei feeddown estimates to be robust with regard to the possible nuclear reactions in the hadronic phase of heavy-ion collisions that were advocated in Refs.~\cite{Oliinychenko:2018ugs,Xu:2018jff,Vovchenko:2019aoz}.

\paragraph*{Summary and outlook.}

Our statistical model calculations reveal significant feeddown corrections from decays of excited nuclei to the final yields of deuterons, tritons, \tHe, and \fHe~produced in relativistic heavy-ion collisions, in particular at $\sqrt{s_{\rm NN}} \lesssim 10$~GeV, where feeddown contributions are of a similar magnitude as the primordial yields. 
The presence of light nuclei leads to a considerable influence on the equation of state of hadronic matter, both the well-known freeze-out criterion $E/N = 1$~GeV as well as the isentropic trajectories are affected appreciably in the baryon-rich region.

The inclusion of excited states does not universally improve the thermal model description of the available experimental data: while the description of \tHe~and \fHe~ yields at the lowest collision energies is improved, that of deuterons and of \tHe~at NA49 energies is not.
Nevertheless, our results do suggest that the excited nuclei feeddown cannot be neglected in the ongoing and future analysis of light nuclei production at intermediate collision energies.
The feeddown should also be considered in other theoretical approaches to the production of loosely-bound states, such as coalescence or transport theory. 

The future, with the expected results from the beam energy scan programs at RHIC and SPS, results from the HADES experiment, the upcoming high-luminosity measurements at the LHC and the future experiments CBM at FAIR and BM@N at NICA, offer the unique possibility for high precision studies of light nuclei formation all the way from a regime where they are an extremely rare probes  to where they contribute to the bulk of the created matter.


\begin{acknowledgments}

We thank P.~Braun-Munzinger, V.~Koch, D.~Oliinychenko, and J.~Randrup for useful discussions and comments.
V.V. was supported by the
Feodor Lynen program of the Alexander von Humboldt
foundation and by the U.S. Department of Energy, 
Office of Science, Office of Nuclear Physics, under contract number 
DE-AC02-05CH11231.
B.D. acknowledges the support from BMBF through the FSP202 (F\"orderkennzeichen 05P15RFCA1).
H.St. acknowledges the support through the Judah M. Eisenberg Laureatus Chair by Goethe University  and the Walter Greiner Gesellschaft, Frankfurt.

\end{acknowledgments}

\bibliography{excited-nuclei}


\end{document}